%% file: ijcai25.tex
\documentclass{article}
\usepackage{ijcai25}

\usepackage{times}
\usepackage{soul}
\usepackage{url}
\usepackage[hidelinks]{hyperref}
\usepackage[utf8]{inputenc}
\usepackage[small]{caption}
\usepackage{graphicx}
\usepackage{amsmath}
\usepackage{amsthm}
\usepackage{booktabs}
\usepackage{algorithm}
\usepackage{algorithmic}
\usepackage[switch]{lineno}
\usepackage{cleveref}
\usepackage{tikz}
\usetikzlibrary{calc}
\usepackage{setspace}
\usepackage{enumitem}
\usepackage{siunitx}
\usepackage{xcolor}
\definecolor{lightblue}{RGB}{201, 217, 240}

\title{LeRAAT: LLM-Enabled Real-Time Aviation Advisory Tool}

\author{
Marc R. Schlichting\and
Vale Rasmussen\and
Heba Alazzeh\and
Houjun Liu\and
Kiana Jafari\and \\
Amelia F. Hardy\and
Dylan M. Asmar\And
Mykel J. Kochenderfer\\
\affiliations
Stanford University\\
\emails
\{mschl, valer22, alazzeh, houjun, kjafari, ahardy, asmar, mykel\}@stanford.edu
}

\begin{document}

\maketitle
\begin{abstract}

In aviation emergencies, high-stakes decisions must be made in an instant. Pilots rely on quick access to precise, context-specific information---an area where emerging tools like large language models (LLMs) show promise in providing critical support. This paper introduces LeRAAT, a framework that integrates LLMs with the X-Plane flight simulator to deliver real-time, context-aware pilot assistance. The system uses live flight data, weather conditions, and aircraft documentation to generate recommendations aligned with aviation best practices and tailored to the particular situation. It employs a Retrieval-Augmented Generation (RAG) pipeline that extracts and synthesizes information from aircraft type-specific manuals, including performance specifications and emergency procedures, as well as aviation regulatory materials, such as FAA directives and standard operating procedures. We showcase the framework in both a virtual reality and traditional on-screen simulation, supporting a wide range of research applications such as pilot training, human factors research, and operational decision support. The code\footnote{\url{https://github.com/sisl/LeRAAT}} and a demo video\footnote{\url{https://youtu.be/NnijQAlTo-U}} are publicly available.
\end{abstract}

\section{Introduction}
\begin{figure}[t]
    \centering
    \input{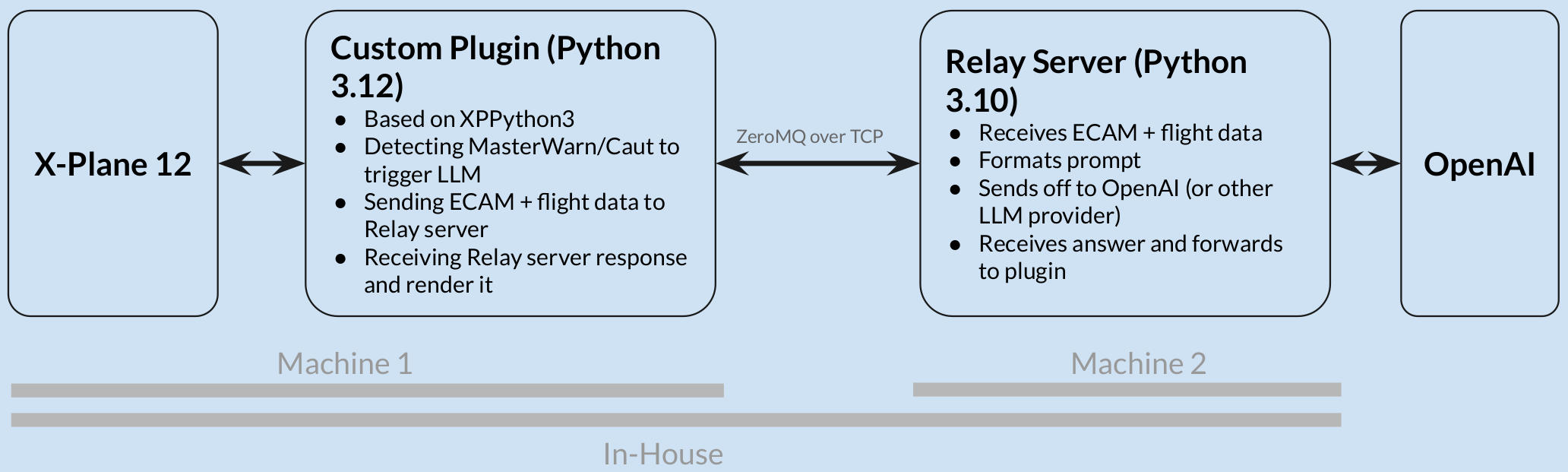}
    \caption{LeRAAT system architecture. Highlighted in blue are our main contributions.}
    \label{fig:system}
\end{figure}
Modern aviation has an exceptional safety record with accident rates of one per 1.6 million flights \cite{osz:aviation-safety}. However, in 2024 alone, 268 people lost their lives in commercial aviation accidents \cite{ASN24}. A recent study found that human factors contribute to approximately 75\% of  aviation accidents and incidents, with the most prevalent human factor being the loss of situational awareness in the cockpit \cite{kharoufah2018review}. While existing aviation systems effectively maintain basic safety envelopes, there remains a gap in systems capable of emulating the sophisticated aeronautical decision-making and contextual judgment of an experienced pilot \cite{NASA2018considerations}.

Recent advances in large language models (LLMs) offer a promising solution to this challenge given their ability to process complex information streams and deliver contextually relevant recommendations \cite{zhao2023:survey}. Their capability to synthesize multiple data sources makes them potentially valuable tools for pilot decision support \cite{bommasani2022opportunitiesrisksfoundationmodels}, such as recommending a suitable alternate airport based on specific system failures and current weather conditions.

This paper presents LeRAAT, a framework that integrates LLMs with the X-Plane flight simulator to explore the potential of AI-assisted pilot decision support. Our system combines real-time flight data, weather information, and aircraft documentation through a Retrieval-Augmented Generation (RAG) \cite{10.5555/3495724.3496517} architecture  to provide contextually relevant recommendation while maintaining pilot authority. The primary contributions of this work include: 
(1) a modular architecture that enables seamless integration of LLMs into flight simulations in X-Plane,
(2) a RAG implementation incorporating aircraft manuals and procedures and, (3) a pilot-facing interface, available as an overlay window within the simulator and compatible with virtual reality (VR), designed to minimize interference with pilot workflow.

Through an initial evaluation with professional pilots, we explore the system's ability to provide relevant recommendations while maintaining pilot authority. To illustrate its functionality in real-time flight scenarios, we present a demonstration featuring both VR and non-VR cockpit interfaces, showcasing how the system adapts to different flight situations. 
Our preliminary findings represent early steps toward understanding whether and how LLMs could be integrated into the cockpit.

\section{System Architecture}
Our framework employs a modular architecture, as shown in \cref{fig:system}. The framework consists of three main components: (1) the user interface, which is integrated into X-Plane as a plugin, (2) a relay server for data processing, prompt assembly, routing, and (3) an LLM backend for response generation. This modular design ensures system flexibility by allowing easy substitution of the LLM backend, adaptation to different aircraft types through documentation swapping, and decoupling of user interface and backend feature development.

\begin{figure}[t]
    \centering
    \includegraphics[width=\linewidth]{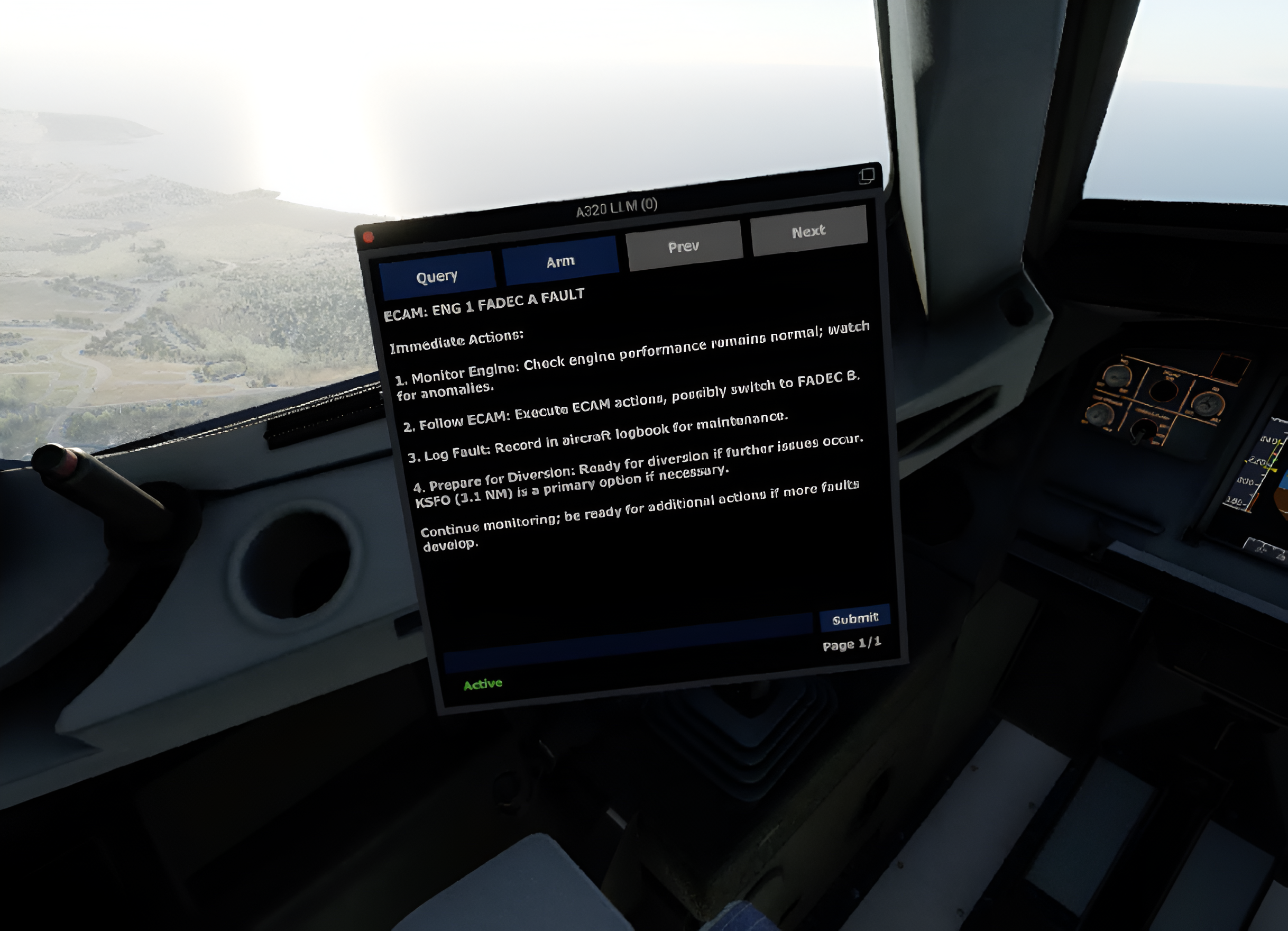}
    \caption{Graphical user interface with LLM response to a full authority digital engine control (FADEC) failure.}
    \label{fig:user_interface}
\end{figure}

\paragraph{ X-Plane Integration.}
The system interfaces with X-Plane through \texttt{XPPython3}\footnote{\url{https://xppython3.readthedocs.io/}}, a wrapper around the official X-Plane SDK. This enables real-time data extraction and interactive display capabilities within the simulator environment. While our framework is aircraft agnostic, we use the Airbus A320NEO (Toliss) as a representative example. X-Plane and the Toliss A320NEO offer high-fidelity simulation of both flight dynamics and aircraft systems. Our plugin accesses live aircraft data including electronic centralized aircraft monitor (ECAM) messages, autothrottle modes, and autopilot states. These parameters provide essential context for the AI assistant.
\paragraph{Interactive UI Components.}
The user interacts with the LLM assistant through a graphical user interface (GUI) which is displayed as an overlay window on the screen within X-Plane or in a fixed position in the cockpit if VR is used. An example of the user interface in VR is shown in \cref{fig:user_interface}. Important features of the user interface include font size selection to guarantee legibility in VR and pagination for longer LLM outputs. The overall design of the user interface takes inspiration from Airbus' \textit{dark cockpit} philosophy, which emphasizes a reduction in unnecessary alerts to minimize pilot workload 
\cite{AirbusDesign}.

\begin{figure}
    \centering
    \resizebox{\columnwidth}{!}{
    \begin{tikzpicture}
        \node[draw,thick,rounded corners, minimum width=3cm, align=center, fill=black!10, inner sep=1em] (armed) at (0,0) {{\large \textbf{Armed}} \\ blank screen};
        \node[draw,thick,rounded corners, minimum width=3cm, align=center, fill=black!10, inner sep=1em] (active) at (8,0) {{\large \textbf{Active}} \\ display LLM text};
        \node[draw,thick,rounded corners, minimum width=3cm, align=center, fill=black!10, inner sep=1em] (interactive) at (4,5.5) {{\large \textbf{Interactive}} \\ interactive mode};
    
        \draw[-stealth, thick] (armed) to[out=55, in=125, looseness=0.7] node[midway,above,align=center] {activate through \\ \textit{Query} button} (active);
        \draw[-stealth, thick] (armed.east) to[out=20, in=160] node[midway,above,align=center] {activate through \\ Master Caut/Warn} (active.west);
        \draw[-stealth, thick] (active) to[out=235, in=-55, looseness=0.7] node[midway,above,align=center] {\textit{Arm} button} (armed);
        \draw[-stealth, thick] (armed) to[out=110, in=170] node[midway, left=0.5em, align=right] {text box entry \\ \textit{Submit} button} (interactive);
        \draw[-stealth, thick] (interactive) to[out=-120, in=80] node[pos=0.5, above=0.5em, align=center, rotate=45] {text box entry \\ \textit{Submit} button} (armed);
        \draw[-stealth, thick] (active) to[out=100, in=-70] node[pos=0.35, above=0.4em, align=center, rotate=-37] {text box entry \\ \textit{Submit} button} (interactive);
        \draw[-stealth, thick] (interactive) to[out=10, in=70, looseness=1.2] node[pos=0.5, right=0.5em, align=left, rotate=0] {\textit{Query} button} (active);
        \draw[-stealth, thick] (active.north east) to[out=45, in=-45, looseness=3] node[midway,right,align=center] {update through \\ \textit{Query} button} (active.south east);   
    \end{tikzpicture}
    }
    \caption{Discrete state model of LeRAAT.}
    \label{fig:system_states}
\end{figure}
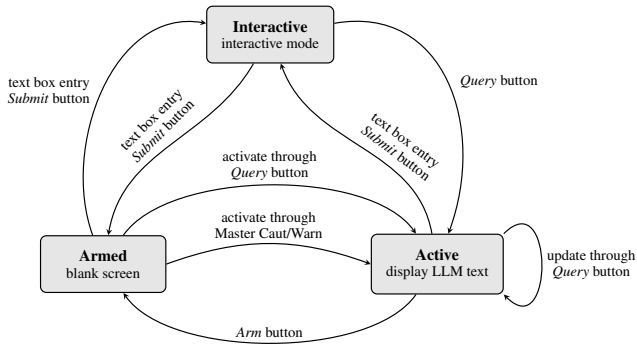

The system operates in three discrete states as depicted in \cref{fig:system_states}: \textit{Armed} for passive monitoring of aircraft warnings and cautions, \textit{Active} for a triggered response mode that provides situation-specific guidance, and \textit{Interactive} enabling direct pilot-LLM conversation. The transitions between the states are shown in \cref{fig:system_states}.

\paragraph{Relay Server.}
The relay server facilitates communication between the plugin and the AI assistant. When triggered by a pilot query, master warning, or master caution, our plugin sends a comprehensive set of flight parameters and all ECAM messages to the relay server. Within the relay server, this data undergoes preprocessing, is used for document retrieval, and assembly of the final LLM prompt. On the return path from the LLM, the response is transmitted through the relay server to the user interface.

\paragraph{LLM Backend.}
The system utilizes GPT-4o as its LLM, though the architecture supports easy transition to other LLM providers, or local LLM instances. 
To accommodate stateless LLMs, the relay server maintains a conversation context, ensuring continuity in interactions with the pilot.

\paragraph{Data Flow and Processing.}
The preprocessing phase consists of three main components: (1) formatting flight data and ECAM messages, (2) retrieving relevant documentation based on flight conditions and ECAM messages, and (3) identifying available runways and weather conditions at close-by alternate airports. 

For the integration of technical documentation we use RAG to extract relevant sections from operating manuals, incorporating reference materials without requiring complete document inclusion.  This approach enhances the quality of recommendations, as RAG enables the use of aircraft-specific, non-public data.

The retrieved document chunks are concatenated with the formatted flight data, ECAM messages, and alternate airport information, and subsequently sent to the LLM. The LLM, generates responses based on the retrieved content. The system prompt directs the LLM to provide concise responses.

\paragraph{Retrieval-Augmented Generation.} \label{sec:rag}
Large language models encode a vast amount of common-sense knowledge \cite{NEURIPS2023_65a39213}, including basic aviation principles and cause-effect reasoning.  However, they have limited capacity as standalone knowledge stores \cite{petroni_language_2019} and can generate incorrect information without grounding.  The \textit{hallucination} effect \cite{xu_hallucination_2024} is difficult to identify without a separate verification mechanism---obviating the purpose of using the LM as a knowledge store. To mitigate this, we employ RAG \cite{10.5555/3495724.3496517}, which retrieves relevant text from an external corpus before generating responses. Broadly, RAG systems use two LLMs working in conjunction to fulfill a user query: a \textit{retrieval} encoder LLM to identify semantically relevant text from a corpus usually trained with contrastive learning, and a \textit{generation} decoder LLM to rephrase retrieved content to fulfill the user query \cite{10.1145/3626772.3657957}.

Aircraft documentation is primarily stored in PDF format. We use \texttt{pymupdf4llm} \cite{pymupdf4llm} 
to convert content into structured text. The extracted text is then divided into smaller, overlapping chunks. Each chunk is embedded using \texttt{text-embedding-3} from OpenAI, and stored in a vector database using Facebook AI similarity search (FAISS) \cite{douze2024faiss}. During a user query, the system retrieves the 10 most relevant document chunks by comparing the embedded query with the stored vectors.

\section{Example Scenarios and Pilot Feedback}
To provide a clearer understanding of how our system works, we will demonstrate how LeRAAT responds to two X-Plane flight scenarios.

In the first scenario, a pilot is operating an Airbus A320NEO at flight level \num{320} on a routine flight from Miami to San Francisco when they notice a concerning fuel imbalance between the left and right wing tanks. Upon activating the AI assistant through the GUI, the system immediately receives the current flight parameters, weather, and nearby airports. The assistant identifies a potential fuel leak and provides a prioritized list of diversion options based on runway lengths, current weather conditions, and distances to nearby airports. During this interaction, the pilot engages with the AI assistant through the \textit{Interactive} mode to ask specific questions about the situation, such as ``What's our current fuel endurance?''

In the second scenario, an A320NEO is on final approach to Seattle-Tacoma International Airport in poor weather when they receive a master warning and an indication that two hydraulic pumps have failed. The advisory system is automatically activated and receives ECAM messages about the failures. After analyzing the situation, LeRAAT advises the pilot to proceed with the landing, despite some degradation in flight controls and braking systems, due to the expectation of further system failures with prolonged flight time.

In fact, these two particular scenarios were used to evaluate the usefulness of our system. We invited three professional pilots to fly these simulated scenarios and, as subject matter experts (SMEs), provide feedback on the recommendations given by the system.
All three pilots emphasized the LLM system’s unique ability to evaluate nearby airports and recommend options to divert. One of our experts, a current A320 pilot, said “providing synthesis of where you are, where you can go, and for which reasons... is something that’s missing in other systems but was provided here.” While not necessarily a measure of success, we also noticed that pilots who followed the LLM system’s instructions were able to resolve the situation more quickly than those who did not. This was especially true for the second scenario, where a quick decision must be made to continue or abort the approach.

SMEs unanimously encouraged switching the system to use voice communication instead of text input. Supplying the LLM system with more information about runway directions and notices to air missions (NOTAMs) was another area of improvement highlighted by our SMEs. This feedback from SMEs helped identify the key benefit of our system: a pilot can receive information tailored to their exact situation and can continue receiving guidance until the aircraft is safely stopped. Our demo video features two additional scenarios where LeRAAT demonstrates these capabilities.

\section{Conclusion and Future Work}
We present LeRAAT, a framework that integrates LLMs with real-time flight simulation to explore AI-assisted pilot decision support, especially in emergency situations. By combining live aircraft data, weather conditions, and aircraft manuals through a RAG architecture, the system provides pilots with context-aware recommendations tailored to their specific flight situation.

As an initial proof-of-concept, we evaluated the system in simulated flight scenarios and gathered feedback from professional pilots. Their responses highlighted the potential of AI-generated advisories in improving situational awareness and providing structured decision support. However, challenges remain, such as the integration of additional flight data sources such as NOTAMs, and the need for a voice-based interaction system to streamline pilot-AI communication. 

The modular design of our framework enables easy adaptation to specific research questions, allowing the broader research community to explore various aspects of LLMs in the cockpit.
Based on pilot feedback, we expect LLMs to provide the greatest utility in general aviation operations, where pilots typically have less extensive training compared to commercial aviation. Our research team is currently conducting a comprehensive quantitative study of the LeRAAT framework within the general aviation domain.

\section*{Ethical Statement}
Aviation is a safety-critical domain, where erroneous or delayed recommendations can have serious consequences. While LLMs enable dynamic advisory generation, they can also “hallucinate” or produce contextually incorrect suggestions. To mitigate these risks, we employ validation methods at multiple stages:

\textbf{Human-in-the-Loop Review:} Pilots and subject-matter experts remain the final authority, using LLM outputs as supplemental guidance rather than absolute instructions.

\textbf{Restricted Prompting and Guardrails:} We use domain-specific prompts and scenario constraints to reduce the likelihood of irrelevant or hazardous recommendations.

These measures help ensure that LLM-driven advisories do not compromise safety. Ongoing work includes refining prompt engineering techniques, incorporating confidence metrics in real time, and extending system validation with additional real-world pilot feedback.

\section*{Acknowledgments}
We would like to thank Acubed for their generous support and insights throughout the project.

\section*{Authorship and Responsibility}
All human authors listed take full responsibility for the accuracy, originality, and integrity of this work. Every piece of analysis, design decision, and final written content was reviewed, verified, and approved by the authors. We confirm that any external contributions are duly acknowledged, and we hold ourselves accountable for any content presented in this paper.

\bibliographystyle{named}
\bibliography{ijcai25}

\end{document}

%% file: images/system_architecture.tex

\begin{tikzpicture}[
]


\node[draw,rounded corners, thick, minimum width=8cm, align=center, fill=black!15!white] (flightsim) at (0,2.5) {%
    \parbox{5.5cm}{%
    \raggedright
    \textbf{Flight Simulator (X-Plane)}
    }
    \parbox{1.5cm}{%
    \includegraphics[width=1cm]{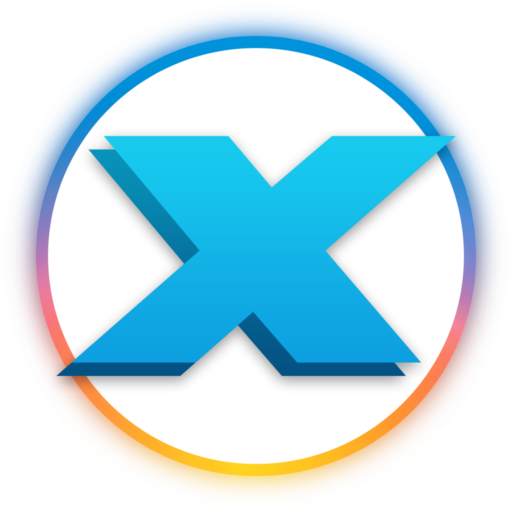}
    }
    
};

\node[draw,rounded corners, thick, minimum width=8cm, align=center, fill=lightblue] (simulator) at (0,0) {%
    \parbox{5cm}{%
    \raggedright
    \textbf{User Interface} \\
    {\footnotesize \begin{itemize}[topsep=0pt, partopsep=0pt, itemsep=0pt, parsep=0pt, left=2pt]
        \item listens to master warning/caution
        \item sends flight data to relay server 
        \item displays recommendation to pilots 
        \item interface for interactive communication
    \end{itemize}}
    }
    \parbox{2cm}{%
    \includegraphics[width=2cm]{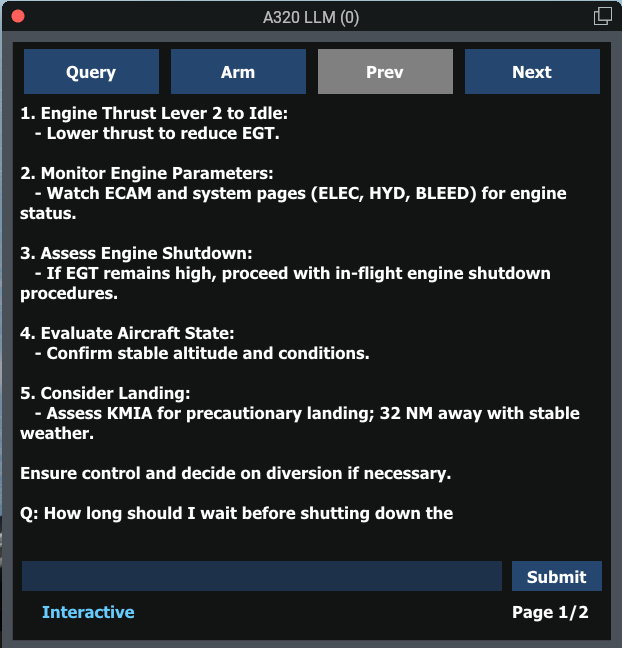}
    }
    
};

\node[draw,rounded corners, thick, minimum width=7cm, align=center, fill=lightblue] (relay) at (-0.5cm,-3.3cm) {%
    \parbox{4cm}{%
    \raggedright
    \textbf{Relay Server} \\
    {\footnotesize \begin{itemize}[topsep=0pt, partopsep=0pt, itemsep=0pt, parsep=0pt, left=2pt]
        \item flight data processing
        \item document retrieval
        \item alternate airport selection
        \item prompt assembly
    \end{itemize}}
    }
    \parbox{2cm}{%
    \includegraphics[width=2cm]{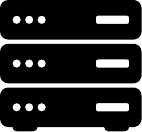}
    }
    
};

\node[draw,rounded corners, thick, minimum width=8cm, align=center,fill=black!15!white] (llm) at (0,-6.4cm) {%
    \parbox{5cm}{%
    \raggedright
    \textbf{LLM} \\
    {\footnotesize \begin{itemize}[topsep=0pt, partopsep=0pt, itemsep=0pt, parsep=0pt, left=2pt]
        \item generates concise response
        \item can easily be swapped out
    \end{itemize}}
    }
    \parbox{2cm}{%
    \includegraphics[width=2cm]{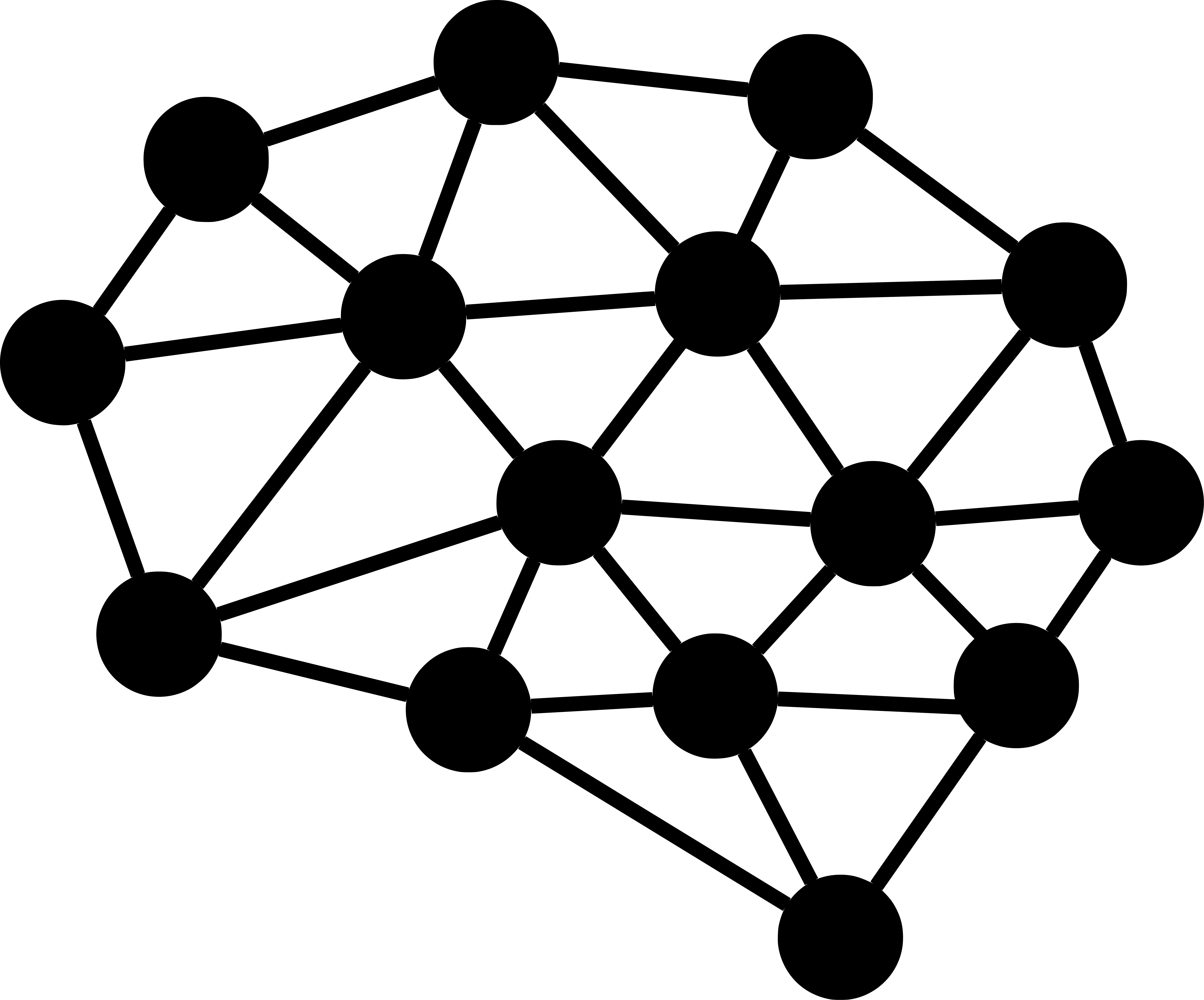}
    }
    
};

\draw[thick, -stealth] ($(simulator.south)+(-1.15,0)$) to[in=135,out=-135] node[pos=0.5, left]{\scriptsize flight data} ($(relay.north)+(-0.65,0)$);
\draw[thick, -stealth] ($(relay.north)+(+0.65,0)$) to[in=-45,out=45] node[pos=0.5, right]{\scriptsize response} ($(simulator.south)+(0.15,0)$);

\draw[thick, -stealth] ($(relay.south)+(-0.65,0)$) to[in=135,out=-135] node[pos=0.5, left]{\scriptsize formatted prompt} ($(llm.north)+(-1.15,0)$);
\draw[thick, -stealth] ($(llm.north)+(0.15,0)$) to[in=-45,out=45] node[pos=0.5, right]{\scriptsize response} ($(relay.south)+(+0.65,0)$);

\draw[thick, -stealth] (flightsim) to[in=90,out=-90] node[pos=0.5, right]{\scriptsize aircraft state} (simulator);

\draw[thick, -stealth] ($(relay.east)+(1,0)$)  -- ($(relay.east)+(0,0)$)
    node[pos=1.0, above=1.3em, align=left, anchor=west]  {\scriptsize aircraft}
    node[pos=1.0, above=0.5em, align=left, anchor=west] {\scriptsize documents} ;

\draw[thick, -stealth] ($(relay.east)+(1,0.8)$)  -- ($(relay.east)+(0,0.8)$)
    node[pos=1.0, above=0.5em, align=left, anchor=west]  {\scriptsize weather} ;

\draw[thick, -stealth] ($(relay.east)+(1,-0.8)$)  -- ($(relay.east)+(0,-0.8)$)
    node[pos=1.0, above=1.3em, align=left, anchor=west]  {\scriptsize airport}
    node[pos=1.0, above=0.5em, align=left, anchor=west] {\scriptsize database} ;

\end{tikzpicture}